# *Relining or Replacement: How Timing, Costs, and Infrastructure Shape U.S. Primary Steel Decarbonization*

Mohamed Atouife[1,2,*], Yuezhang He[3], and Jesse D. Jenkins[1,3,*]

[1] Mechanical and Aerospace Engineering Department, Princeton University, Princeton, New Jersey, 08540, United States of America.

[3] Andlinger Center for Energy and the Environment, Princeton University, Princeton, NJ, United States of America

[2] Lead contact.

[*] Correspondence: m.atouife@princeton.edu, yh9774@princeton.edu, jdj2@princeton.edu

# Summary

Imminent blast furnace relinings create a consequential investment window for U.S. primary steel decarbonization. Using plant-level capacity-planning and stochastic operations modeling, we quantify the costs, infrastructure, and fiscal requirements for transitioning legacy blast furnace sites to direct reduction under supply-side incentives. Under baseline assumptions, post-subsidy premiums above average blast furnace costs, including relining, are $272–427 per ton of steel for near-zero-emission hydrogen-based direct reduction and $115–181 for natural-gas-based production. Representative facilities producing two million tons of steel annually at legacy locations require approximately 4–6.5 and 0.7–1.1 GW of renewable generation and battery storage, respectively. This translates to 65 and 12 GW across the unrelined fleet for hydrogen and natural gas-based pathways respectively. These findings connect replacement timing to costs, infrastructure, and public support, motivating a sequenced transition through clean-powered natural-gas-based direct reduction to avert blast furnace relining, reduce emissions by 57% relative to blast furnaces, and preserve options for deeper decarbonization through hydrogen substitution or carbon capture.

# Introduction

The U.S. primary steel sector faces a concentrated set of reinvestment decisions that will shape its production technologies and emissions well into the 2040s[1]. Before 2030, at least nine of the country's seventeen blast furnaces face relining decisions, affecting approximately 15 million tons per year (mmt/yr) of ironmaking capacity. Including mothballed furnaces brings the capacity potentially affected to approximately 22.5 mmt/yr, out of a total of 32 mmt/yr across eight integrated steelmaking sites. The remaining 9.5 mmt/yr has already been relined, underscoring that this investment cycle is underway. Relining can extend blast furnace operation by up to twenty years[2], potentially sustaining coal-based production through another capital cycle. Across announced relinings, open decisions, and mothballed capacity, approximately 644 $MtCO_2$ of emissions are at stake over a twenty-year operating horizon (Figure 1). This investment cycle presents a distinct, policy-relevant opportunity to determine the emission trajectory of domestic primary ironmaking.

Although roughly 70 percent of U.S. steel output already comes from scrap-fed electric arc furnaces (EAF)[3], scrap alone cannot fully substitute for primary production on the investment timelines implied by relining. Scrap supply is limited due to high recycling rates (reaching 90% and 100% for construction and the automotive sector[4]). Additionally, primary iron is needed to dilute impurities and scrap quality limits driven by the accumulation of tramp elements to meet demanding product specifications (e.g. for automotive steel)[5–7]. The central decarbonization challenge for primary steel in the United States is therefore immediate, namely, what replaces coal-based ironmaking at incumbent sites as assets turn over.

Direct reduced iron (DRI) paired with EAFs provides an alternative production route using natural gas, hydrogen, or a combination of reductants, with carbon capture and storage (CCS) offering an additional abatement option. A substantial literature evaluates these configurations through techno-economic analysis, energy-system modeling, and comparisons of emissions performance. Studies identify hydrogen-based DRI–EAF as a promising near-zero-emission pathway whose competitiveness improves with lower renewable electricity costs, lower electrolyzer capital costs, and carbon pricing[8–13]. Moreover, spatial assessments have identified key geographies such as Australia, where the co-location of high-quality iron ore and renewable energy potential creates favorable conditions for green steel production[14]. Parallel research examines CCS at integrated blast furnace–basic oxygen furnace (BF–BOF) facilities, where dispersed emissions sources and plant integration complicate capture[15,16], and at natural-gas-based DRI facilities, where concentrated process $CO_2$ streams are comparatively straightforward to capture[12,16].

The literature underscores the importance of phasing out blast furnaces[17], and establishes the technological and economic conditions under which lower-carbon steelmaking can compete, but comparisons of production costs and favorable locations do not, by themselves, establish how replacement can occur within an existing industrial system. Steel assets are long-lived and embedded in established workforces, supply chains, and infrastructure. Their replacement proceeds through discrete reinvestment decisions at legacy sites rather than unconstrained development on a blank slate. The relevant question is therefore not whether a pathway can reach parity in an abstract least-cost setting, but whether it is feasible under timelines dictated by asset turnover and existing policy to transition existing sites to lower-carbon production or to phase these sites out in favor of greenfield replacements. This places the transition at the intersection of timing, technology, and policy: asset turnover establishes the investment window, technology choice determines the supporting infrastructure required, and policy support changes the economics of undertaking that investment.

The central issue is how the technologies and policy instruments available at a particular investment window shape the options for industrial transition. The Inflation Reduction Act of 2022[18] (IRA) provides a test of a supply-side approach to this challenge: whether subsidizing clean hydrogen, clean electricity, and carbon capture could narrow production-cost gaps sufficiently to redirect reinvestment toward lower-carbon steelmaking during the current relining cycle. Although several incentives were later weakened, we assess the extent to which the original framework could have made these investments economically viable.. This paper asks two questions. First, under legacy-site conditions, would the IRA's supply-side credits have been sufficient to unlock large-scale investment in lower-carbon primary steelmaking in the United States, and at what fiscal cost when scaled across the U.S. fleet? Second, where the IRA tax credit architecture falls short, what do those shortfalls reveal about the design of more effective industrial decarbonization policy in the future and in other jurisdictions?

To this end, we evaluate the production costs, emissions, supporting infrastructure, and fiscal requirements of alternative DRI–EAF configurations at incumbent U.S. steelmaking locations. This study develops a stochastic optimization model of the steelmaking process that accounts for variability in wind and solar across different weather years and considers the impact of production and investment tax credits (PTC and ITC) as well as carbon capture credits established by the Inflation Reduction Act. The model leverages six years of hourly weather data to optimize the capacity needed for DRI production, electric arc furnaces, electrolysis when relevant, and additional renewable energy technologies such as wind, solar, and battery storage, and uses results to estimate location-specific economic factors, including the expected levelized cost of steel (LCOS), electricity, and hydrogen from water electrolysis. The study considers both grid-connected and fully islanded operational configurations, subject to a $CO_2$ emissions limit relative to BF-BOF emissions. In the islanded configuration, the facility operates exclusively on dedicated renewable energy resources (complemented by energy storage), a strategy that can circumvent the extensive interconnection queues and delays commonly encountered in the United States[19]. For hydrogen-based pathways, the modeled grid-connected facility differentiates between hydrogen produced using grid electricity and that generated from dedicated renewable resources, ensuring that the hydrogen production tax credit applies exclusively to the latter. For both configurations, the stacking of clean electricity and, when relevant, clean hydrogen incentives is essential for reducing the cost of clean electricity and hydrogen production and, by extension, enhancing the economic viability of low-carbon steel manufacturing. A full description of our modeling approach and experimental design is available in the Methods section.

# Results

## Supply side incentives narrow parity gaps but does not eliminate them

Our modeling results indicate that the IRA's supply-side incentives for inputs to lower-carbon steelmaking reduce but do not eliminate cost gaps between incumbent BF–BOF steelmaking and low-carbon DRI–EAF pathways at legacy sites. Residual premiums vary substantially across pathways and geographies, and these differences translate into sharply different infrastructure requirements and fiscal exposure.

Throughout this section, production-cost premiums are measured relative to average BF–BOF benchmark levelized costs of steel (LCOS) which spans $420–529 per ton of steel, including annuitized relining costs.

As shown in Figure 2, under baseline technology and financing assumptions, unsubsidized $H_2$-DRI–EAF costs range from $950 to $1,117 per ton of steel across legacy locations. Credits for hydrogen and electricity supply reduce these costs by $203–215 per ton of steel, leaving premiums of $272–427 per ton of steel above average BF–BOF costs. Corresponding post-IRA premiums are $115–181 per ton of steel for NG-DRI–EAF, $138–223 per ton of steel for NG-DRI–EAF with CCS, and $210–332 per ton of steel

for the hybrid NG–$H_2$ configuration. Delivered natural gas prices materially affect the NG-based results. Replacing baseline industrial prices of $6–9/MMBtu with the $2–4/MMBtu power-sector price range considered in the sensitivity analysis lowers NG-DRI–EAF costs by up to $80 per ton of steel. At $3/MMBtu, a premium of $77–112 per ton of steel remains above average BF–BOF costs (each additional $1/MMBtu increases LCOS by approximately $12 per ton of steel).

The IRA lacks a specific, dedicated policy mechanism for decarbonizing steelmaking, instead subsidizing clean electricity, hydrogen and CCS more generally, irrespective of applications. The differences in support reflect the inputs eligible for credits under each pathway. For $H_2$-DRI–EAF, approximately $100 per ton of steel of the LCOS reduction comes from the 45V hydrogen credit, with the remainder attributable to renewable electricity and battery storage incentives. The hybrid configuration receives $97–104 per ton of steel in support, reflecting its lower hydrogen requirement. NG-DRI–EAF with CCS receives $73–83 per ton of steel, including approximately $40 per ton of steel for carbon capture and storage, whereas NG-DRI–EAF receives $19–21 per ton of steel entirely through electricity-supply incentives (see supplementary information Figures 2 to 6).

These comparisons also involve different abatement depths (see Figure 2). The baseline $H_2$-DRI–EAF configuration approaches near-zero emissions, whereas NG-DRI–EAF supplied with clean electricity reduces emissions by approximately 57% relative to BF–BOF. We first examine the infrastructure underlying these configurations, then compare hydrogen-based production at a similar emissions reduction to natural gas-based production.

## Renewable resource quality shapes regional cost disparities: contrasting Indiana and Pennsylvania

The range of production and subsidy costs described above reflect substantial geographic variation across existing BF-BOF steelmaking sites in the United States driven by differences in local renewable resource quality. Under IRA incentives, near-zero emission $H_2$-DRI–EAF is substantially more competitive in Indiana than in Pennsylvania, with an LCOS difference exceeding $150 per ton of steel. This contrast suggests that H2-DRI-EAF may not be a universal decarbonization solution for all legacy sites. This large cost advantage stems from differences in the clean electricity build required to satisfy the modeled emissions constraint: a representative steelmaking facility with 2 mmt/yr output requires on the order of 4 GW of dedicated renewable generation and storage in Indiana versus 6.5 GW in Pennsylvania, reflecting both the sheer scale of power infrastructure needed to deliver clean power for hydrogen production (which represents ~70% of total load) and electric arc operations operations, and the variation in local resource quality (with Indiana exhibiting higher quality wind/solar resources).

To isolate the role of renewable resource quality from legacy-site constraints, we evaluate a greenfield $H_2$-DRI–EAF facility in renewable resource-rich Oklahoma under the same stringent emissions constraint. Relative to Pennsylvania, LCOS is $284 per ton lower in Oklahoma, as required dedicated renewable capacity declines from 6.5 GW to 3.4 GW (for a 2 mmt/yr facility). Despite this improvement, the modeled post-IRA premium remains high ($191 per ton relative to average BF–BOF costs), indicating that even high-quality renewable resources do not fully close the parity gap under current technology costs and the IRA's policy incentives.

Replacement is therefore not solely a question of closing a production-cost gap. Particularly for hydrogen, it requires a large infrastructure buildout alongside the steelmaking assets, integrating renewable generation, storage, and reductant production. The scale of this supporting system introduces a deployment consideration distinct from cost competitiveness: steel-asset replacement must be coordinated with the development of the significant energy infrastructure needed to operate it. These requirements are consequently relevant to assessing conversion within the relining window, even where financial support narrows the production-cost premium.

## Cost structure: energy supply dominates $H_2$-DRI–EAF costs

The high $H_2$-DRI–EAF premium reflects a cost structure dominated by upstream electricity and hydrogen supply under a stringent emissions constraint. For near-zero emissions $H_2$-DRI–EAF, electricity and hydrogen infrastructure accounts for 45–53% of LCOS, while ironmaking and steelmaking assets contribute 11–12%. The remainder reflects operating inputs and fixed costs, including iron ore, labor, and consumables. A representative 2 mmt/yr facility requires total capital expenditures ranging from $10 to $14 billion (lower in Indiana, higher in Pennsylvania). Of this total, $2 billion is allocated to electrolysis, with the remaining capital investment dedicated to iron and steel making assets, electricity generation and battery storage.

Excluding blast furnaces already relined in the current investment cycle, replacing the remaining BF–BOF capacity (approximately 25 Mt/yr) with $H_2$-DRI–EAF would require approximately $141 billion in investment under current technology costs. The bulk of this investment, about $115 billion, would be dedicated to financing around 61 GW of dedicated generation and battery storage and 11.7 GW of electrolysis. Only roughly $27 billion would be allocated to the direct reduction and electric arc furnaces (see Figure 3).

In comparison, the clean-powered NG-DRI–EAF route, which yields lower yet substantial emissions reductions of roughly 57% relative to BF–BOF, requires significantly less dedicated clean energy infrastructure, at roughly 0.7–1.1 GW per 2 mmt/yr facility. This translates to a much smaller investment of roughly $52 billion to replace the remaining BF-BOF capacity. For NG-DRI–EAF, the cost distribution is inverted: the majority, approximately $32 billion, is allocated to the core steel assets. The remaining $20 billion is allocated to the buildout of 12 GW of solar, wind, and battery storage assets.

## Grid-assisted operation reduces costs modestly under the modeled 45V constraint

The infrastructure comparison above partly reflects the deeper abatement required of the baseline hydrogen configuration. To evaluate hydrogen-based production at an emissions reduction comparable to clean-powered NG-DRI–EAF, we examine a grid-assisted $H_2$-DRI–EAF configuration constrained to a 60% reduction relative to BF–BOF. In this configuration, the model uses dedicated clean electricity primarily for electrolysis while relying more heavily on grid electricity for EAF and balance-of-plant loads.

This operating pattern is an endogenous response to the policy constraint implemented in the model: 45V is awarded only to hydrogen produced from clean electricity, which creates an incentive to allocate clean power to electrolysis to maximize the credit value while meeting the emissions target. Relative to the dedicated, near-zero-emission hydrogen configuration, grid-assisted operation at the 60% reduction target lowers LCOS by $50–120 per ton of steel across legacy sites, as the installed renewable power capacity falls to 2.3-2.5GW. Production costs nevertheless remain $217–306 per ton of steel above average BF–BOF costs, compared with $115–181 per ton of steel for NG-DRI–EAF at its approximately 57% reduction under baseline gas-price assumptions ( $77-112 per ton assuming $3 per MMbtu natural gas prices). The higher cost of H2-DRI-EAF is therefore not solely a consequence of the near-zero-emission requirement.

## Complementary financial measures and technological change narrow the remaining gaps

Given that the suite of tax credits established by the IRA are insufficient to close the cost premium for low-carbon steelmaking relative to BF-BOF, we explore additional/complementary policies and the role of potential technological change in closing the gap. The substantial capital requirements for steel assets and their supporting energy systems, particularly for hydrogen-based production, motivate three levers that act directly on capital costs: (i) projected 2030 technology costs (these projections include reductions of up to 60 percent for electrolysis, 31 percent for solar photovoltaics, 21 percent for wind power, and 45 percent for battery storage, as detailed in the supplementary information), (ii) a 100-basis-point reduction in the cost of capital (our base-case analysis assumes technology-specific capital costs ranging from approximately 4% for mature technologies, such as solar, to about 7% for electrolyzers, see supplementary information) , and (iii) a hypothetical 30% industrial investment tax credit (ITC) applied to DRI and EAF assets. The industrial ITC is modeled analogously to clean-electricity investment credits, extending investment support to the core steelmaking equipment. Such upfront support could alternatively take the form of a capital expenditure grant.

If they materialize, projected cost declines by 2030 are expected to narrow residual cost premiums across pathways: inclusive of IRA credits, $H_2$-DRI–EAF premiums fall to $90–192 per ton in Indiana and $201–342 per ton in Pennsylvania, while NG-DRI–EAF premiums decline to $62–72 per ton and $88–105 per ton, respectively. As the NG-DRI-EAF pathway consumes less electricity and no hydrogen, the importance of technology cost reductions is relatively modest, ranging from a reduction of $5-15 per ton of steel. Furthermore, across the modeled pathways and locations, reducing the cost of capital by 100 basis points lowers LCOS by $27–37 per ton of steel, while the industrial ITC provides an additional reduction of $20–25 per ton of steel.

The combined optimistic cases include reduced technology costs, lower financing costs, the industrial ITC, and $3/MMBtu delivered natural gas where applicable. $H_2$-DRI–EAF retains premiums of approximately $62 per ton of steel above average BF–BOF costs in Indiana and $162 per ton of steel in Pennsylvania. NG-DRI–EAF premiums decline to $33–57 per ton of steel above average BF–BOF costs, falling within the $420–529 per ton of steel BF–BOF cost range without reaching its average across many states (see Figure 2).

Since these levers do not achieve cost parity, we use the model results to quantify the output-linked support required to close the remaining parity gap. Such steel production tax credits have been floated by congressional bills[20]. Assuming a steel production tax credit applied over 10 years, a parity-calibrated credit under current costs would range from $429-689 per ton for $H_2$-DRI–EAF (lower in Indiana, higher in Pennsylvania), declining to $81-123 per ton under the most optimistic assumptions. For NG-DRI–EAF, the credit value is $124-181 per ton. When combined with an industrial ITC and a 100-basis point reduction in the cost of capital, the necessary production tax credit for NG-DRI-EAF steel would range from $74 to $126 per ton. This required credit could drop further, to between $54 and 92 per ton, as the cost of power technologies decreases.

### Supply-side incentives entail substantial fiscal commitments at scale

Transitioning to low-carbon steelmaking across the unrelined U.S. blast furnace fleet implies substantial tax expenditures under the IRA-like supply side credits, particularly for hydrogen pathways. As shown in Figure 4, fleet-wide transition to $H_2$-DRI–EAF implies approximately $76 billion in federal tax expenditures over the credit window, with roughly half attributable to the 45V hydrogen credit. The corresponding fiscal burden is lower for clean-powered NG-DRI–EAF, at roughly $7.5 billion driven entirely by clean electricity and storage incentives, and for NG-DRI–EAF with CCS, at approximately $28 billion, including about $14 billion from carbon capture and storage credits, with the remainder from clean electricity and battery storage incentives. A 30 percent industrial investment tax credit applied to core steel assets would add approximately $6.6 to 8 billion in fiscal cost across the unrelined blast furnace fleet, equivalent to roughly $500 million for a representative 2 mmt/yr facility. Finally, the fiscal exposure of an output-linked steel production tax credit scales linearly with the credit rate, amounting to approximately $25 billion over ten years for each $100 per ton of credit value.

Expressed per ton of $CO_2$ avoided, expenditure under the modeled IRA incentives is approximately $15 for NG-DRI–EAF and $85 for $H_2$-DRI–EAF (Figure 5). Although $H_2$-DRI–EAF delivers larger absolute lifetime abatement (~886 $MtCO_2$ per plant) than clean-powered NG-DRI–EAF (~487 $MtCO_2$), it does so at higher implied public expenditure per unit of avoided emissions under the modeled policy structure. Lifecycle emissions for NG-based pathways are sensitive to upstream methane leakage; in the baseline case, we assume a 2.3% leakage rate (U.S national average[21]), which contributes ~170 $kgCO_2e$ per ton to NG pathway emissions under GWP100 (each incremental percentage leaking point adds roughly 70 $kgCO_2e$ per ton of steel).

Together, the results identify three dimensions of replacement relevant to the relining window: the remaining parity cost gap, the scale of supporting infrastructure, and the public expenditure associated with deployment.

# Discussion

Decarbonizing primary steel production presents not only a technical challenge but also a strategic one: how to deploy scarce fiscal resources and massive enabling infrastructure in a way that aligns with the timing of real investment cycles within the steelmaking industry. Much of the industrial decarbonization literature evaluates technologies relative to a fully decarbonized end state. In primary steelmaking, hydrogen-based direct reduction frequently emerges as the long-run near-zero emission technology. However, an endpoint focus can abstract from the economic, fiscal, and temporal constraints that govern capital replacement in existing industrial systems, where decarbonization unfolds through episodic reinvestment decisions shaped by incumbent assets, relative input prices, and policy design.

Our results show that when evaluated at legacy blast furnace sites under current U.S. cost conditions, achieving cost-parity for clean electrolytic H2-DRI-EAF steel production remains a distant goal. This low-carbon pathway exhibits significant regional cost variations compared to traditional steelmaking and necessitates considerable financial backing to overcome its high premiums. As existing BF-BOF facilities were sited based on legacy conditions related to coal availability, proximity to industrial manufacturing and other considerations independent of available renewable energy supply, converting U.S. BF-BOF plants to H2-DRI-EAF is more costly than greenfield development in renewables-rich regions. Moreover, the challenges extend beyond just financial constraints. The scale of enabling infrastructure at legacy sites is itself a binding near-term barrier. For example, Indiana hosts roughly twenty million tons of BF–BOF capacity, representing the largest cluster of steelmaking capacity in the U.S. by far. Converting all of this capacity to H2-DRI-EAF would require approximately 40 GW of new renewable power and storage capacity. For context, the state presently has 27 GW of installed capacity (about half of which is coal-based). These requirements for large amounts of new electricity generating capacity are further challenged in a period of massive data-center buildout and intensifying competition over scarce electric supplies. This does not diminish hydrogen's long-run importance, but it does imply that treating $H_2$-DRI-EAF as the singular focus of steel decarbonization risks conflating long-run decarbonization ambition with near-term deployability under capital-cycle and infrastructure constraints.

A capital-turnover lens changes what "feasible" decarbonization looks like. NG-DRI-EAF is materially closer to competitiveness and can deliver substantial near-term emissions reductions (on the order of 57%) when paired with low-carbon electricity, with substantially lower enabling-infrastructure requirements than hydrogen-based routes (This pathway also significantly mitigates air pollution from primary steelmaking, relative to BF-BOF, a notable benefit that remains unexplored in this paper.). In this framing, the relevant policy question is not simply which technology minimizes emissions in a steady state, but which sequence of investments maximizes cumulative emissions reductions over realistic capital cycles.

This sequencing logic also reframes concerns about technological lock-in. A narrow view treats NG-DRI-EAF as entrenching natural gas dependence. A capital-cycle perspective instead asks whether the first-stage investment creates a platform that keeps the door open for deeper decarbonization as learning trajectories evolve and infrastructure constraints relax. Direct reduction is compatible with subsequent hydrogen substitution if the economics of green hydrogen production improve. Due to the concentrated nature of emissions, gas-based DRI can also be more readily paired with carbon capture than BF-BOF facilities, opening another path to deeper decarbonization, depending on availability of $CO_2$ transport and geologic storage infrastructure. In this context, a timely transition to NG-DRI-EAF that avoids blast furnace relining investments not only delivers emission benefits but also effectively constitutes a real option to be exercised in the future against two main uncertainties: the pace and scale of electrolysis and hydrogen cost reductions, and the potential for carbon capture and its supporting infrastructure.

Fiscal exposure is the second constraint that changes how pathways rank under realistic policy conditions. Output-based support and supply side incentives scale with production volume and the residual cost premium. In the paper, this shows up starkly in the implied magnitude of credits required for parity and the resulting ten-year fiscal burdens for fleet-wide adoption. When public budgets are constrained, higher subsidy intensity per ton can limit the scale of deployable abatement, even if the technology is technically capable of deeper reductions at the facility level. A central implication is that "cost parity" is not the only relevant metric; the fiscal cost per ton of emissions avoided can reorder near-term priorities, favoring pathways that deliver substantial reductions at lower subsidy intensity while keeping deep-decarbonization options available, namely NG-DRI-EAF.

These constraints also illuminate why an input subsidy only approach is unlikely to be sufficient on its own. Supply-side tax credits reduce input costs, but they do not reliably solve the investment problem created by large, lumpy capital requirements and the absence of long-term revenue certainty for low-carbon steel. Cost parity does not necessarily translate to the *revenue* certainty required to underwrite capital investment. A broader policy portfolio would be more effective if it complemented supply-side incentives with demand-side mechanisms that address a different bottleneck: market risk and the absence of long-term revenue certainty for low-carbon steel. The key design insight is that demand-side instruments can unlock financing by creating predictable, multi-year willingness to pay for verified low-carbon attributes, especially in sectors where upstream price premia have modest effects on final goods. The automotive sector illustrates this dynamic: integrated BF–BOF facilities supply roughly 60% of their output to automotive manufacturing[22], yet even a $100 per ton premium on steel would raise the average vehicle price by only about $90. The binding constraint is therefore not willingness to pay in principle, but the absence of long-term, guaranteed offtake sufficient to justify large capital outlays. "Clean steel" standards can fill this gap by providing a stable demand signal that de-risks investment. Where standards are not politically feasible, voluntary procurement coalitions and aggregated purchases of environmental attributes associated with low-carbon steelmaking can partially substitute[23–25], provided they are scaled and targeted toward avoiding near-term investment decisions that would otherwise lock in emissions for decades.

The feasibility of a large-scale transition to DRI-based primary steelmaking also depends on raw material supply chains. DR-grade iron ore currently represents only about 3% of global iron ore supply, and could become a bottleneck under rapid, large-scale deployment of direct reduction. The United States is relatively well-positioned, with high-quality ore deposits in Minnesota already supporting domestic DRI operations at scale, suggesting ore availability need not be binding in the near term. However, scaling DRI beyond today's footprint would require expansion of beneficiation and processing capacity to raise iron content to DR-suitable levels. This further underscores the importance of demand-side measures as investment certainty is not only required by steel assets, but also the supporting the supply chain.

Taken together, these findings support a theory of change anchored in sequencing and policy complementarity. Near-term strategy should prioritize avoiding new rounds of blast furnace relining by enabling earlier shifts toward NG-DRI–EAF where feasible, with explicit safeguards to limit methane leakage and preserve credible pathways to deeper abatement through hydrogen substitution or carbon capture as those options mature. At the same time, the analysis suggests that the steel sector is unlikely to be a first mover for electrolyzer deployment: even when hydrogen-based steel is an attractive long-run endpoint, its near-term economics and infrastructure requirements at legacy sites imply that the cost reductions needed for widespread adoption will most plausibly be driven by deployment in other sectors with nearer-term willingness to pay. The sheer scale of hydrogen required for an H2-DRI facility equivalent in scale to contemporary BF-BOF plants also far outstrips the likely magnitude of early electrolysis deployment. In contrast, so long as facilities are designed to accommodate blended reductants, shifting to NG-DRI could create the potential for utilizing smaller quantities of H2 to gradually reduce emissions intensity as clean hydrogen supply becomes available (as explored in the hybrid route modeled in this paper). This also underlines the need to evaluate industrial decarbonization *across* sectors rather than one sector at a time, as technology policy and deployment in one industry can reshape the feasible option set in others, particularly when platform technologies that can be used in different sectors, such as electrolysis, are involved.

Crucially, this sequencing logic clarifies why supply-side incentives alone are not sufficient. Even "lower-cost" transitional routes, such as clean-powered NG-DRI-EAF, often retain a residual premium, and output-linked credits can narrow that gap; but closing a modeled parity gap does not guarantee bankable investment conditions, nor can it reliably scale across a sector without exhausting fiscal space. Demand-side standards and procurement address this binding constraint by providing durable offtake for verified low-carbon attributes, converting falling cost premia into financeable projects while limiting reliance on open-ended public subsidies.

Although this analysis is calibrated to U.S. costs and incentives, the broader insight extends beyond the United States and steelmaking. Indeed, most blast furnaces in the world are expected to undergo relining this decade. In any industrial sector characterized by long-lived capital, uncertain learning trajectories, and heterogeneous abatement options, transition pathways are shaped as much by reinvestment timing, infrastructure scarcity, and fiscal feasibility. Industrial decarbonization is therefore not only a problem of identifying the technically optimal end-state technology; it is equally a problem of managing capital turnover under cost uncertainty and fiscal constraint to maximize cumulative emissions reductions over realistic investment cycles.

# Conclusion

The United States is well-positioned to produce lower-carbon steel, with access to abundant low-cost natural gas, emerging $CO_2$ transport and storage capacity, and a strong renewable energy resource base. Yet this analysis shows that decarbonizing its primary steel sector remains a significant challenge. The IRA's supply-side credit architecture, despite its ambition, was insufficient to bring lower-carbon DRI pathways to cost parity with incumbent BF–BOF production at any legacy site under current technology costs and financing conditions. Closing the gap requires a coordinated instrument mix that extends beyond supply-side subsidies to include demand-side clean steel standards, improved financing mechanisms, and coordinated private procurement directed at the specific capital cycle of blast furnace relining.

More broadly, this paper advances a theory of change for steel decarbonization in a resource-constrained policy environment. The economics and infrastructure needs of $H_2$-DRI-EAF limit its role as an alternative to blast furnace relining during the current investment window. The analysis supports a sequenced, multi-sector strategy in which NG-DRI-EAF is deployed at legacy BF–BOF sites in the near term, delivering meaningful emissions reductions of approximately 57% and averting the lock-in of coal-based infrastructure for another two decades. In parallel, electrolysis could be scaled in sectors where hydrogen economics are currently more favorable, generating the learning dynamics and cost reductions that might make blending hydrogen with natural gas or hydrogen-based steelmaking viable over the longer term. Carbon capture, meanwhile, could be advanced through large-scale demonstration projects in direct reduction facilities, preserving a second independent pathway to deep decarbonization. This sequenced approach could maximize emissions reduction leverage per public dollar and aligns industrial investment decisions with realistic technology cost trajectories.

Testing this theory of change rigorously requires analytical tools that go beyond the plant-level modeling conducted here. The renewable electricity requirements of a large-scale transition to DRI, whether hydrogen or natural gas-based, are substantial enough to interact materially with electricity sector investment and dispatch decisions. The next generation of macro-energy systems models, integrating steel sector investment and operations with economy-wide electricity and hydrogen infrastructure planning at detailed temporal and operational resolution, would provide the analytical foundation needed to evaluate how the sequenced decarbonization strategy proposed here performs under different assumptions about grid decarbonization, cross-sector hydrogen demand, and the pace of electrolysis cost reduction. Such models would also allow analysts to capture the competition for scarce renewable electricity capacity across steel, power, and other industrial sectors, a dynamic that single-sector analyses cannot resolve.

# Methods

### DRI-EAF plant configuration

The DRI-EAF facility consists of three main islands:

- An iron reduction and steel production facility with identical equipment across all process variants - whether hydrogen-based, natural gas-based, or hybrid. This island features a direct reduction shaft furnace, briquetted iron storage, and electric arc furnaces. The shaft furnace is supplied with direct-reduction grade pellets, which have an iron content greater than 67%. These pellets are combined with a reductant, such as hydrogen or syngas produced from natural gas reforming, to generate reduced iron. The hot direct reduced iron produced can be directly fed into electric arc furnaces to make crude steel, or it can be cooled and compressed into iron briquettes, which can later be heated and introduced into the EAF.
- A DRI reductant feed island that may consist of either a collocated electrolysis hydrogen production plant, a natural gas reformer, or both, depending on the process configuration. The hydrogen production plant includes a proton exchange membrane (PEM) electrolyzer system that produces hydrogen that can either be fed directly to the shaft furnace or stored in an underground steel-pipe hydrogen storage system.
- An electricity generation island consisting of a collocated mix of solar, wind, and short-duration lithium-ion battery storage dedicated to the plant, supplemented by grid electricity (if allowed for by the emissions constraint set in the model).

The DRI facility is designed for flexible operation across multiple dimensions. Based on communication with domain experts, the DRI shaft furnace can operate dynamically within 70% to 100% of its nominal capacity, with the ability to adjust its output by up to 3.75% of this nominal capacity on an hourly basis (based on a conversation with an expert from the industry). In contrast, the production of crude steel using an EAF is a batch process that typically takes less than 60 minutes per batch[26]. Thus, the EAF is modeled as a continuous dynamic process that can alter production between 0% to 100% of its nominal capacity on an hourly basis. Similarly, the PEM electrolyzers function as dynamic, flexible electricity consumers, capable of rapid response and adjusting their output from a minimum of 7% of nominal capacity to maximum full capacity within minutes[27]. The underground steel pipe storage is constrained by charge and discharge rates of 18% and a cushion gas requirement set at 9% of its total storage capacity[28].

Further information on process stoichiometry, technology cost, and characteristics is available in the supplementary information.

**Stochastic Optimization Models**

We developed three linear programming investment and stochastic operational optimization models for different variants of the DRI-EAF process. One model uses 100% hydrogen, another uses 100% natural gas (with and without carbon capture and storage), and the third is a hybrid that combines 30% hydrogen and 70% natural gas. The three models optimize capacity sizing for each key component and hourly operations to minimize the expected annuitized investment and operational costs subject to a range of technical and process constraints. Operations are optimized across stochastic scenarios with equal probability that capture six years of plant-specific hourly generation capacity factor data for wind and solar, obtained from the National Renewable Energy Laboratory's (NREL) wind and solar toolkits, covering the years 2007 to 2012[29,30]. Since solar and wind project developers can choose between the production and investment tax credits from the IRA, each model was run for three cases: assuming no IRA incentives, assuming PTC, and assuming ITC. See supporting information Section 2 for a complete mathematical formulation of this model.

# Resource Availability

# Lead Contact

Lead contact Inquiries regarding the data and method associated with this paper can be directed to the lead contact, Mohamed Atouife (m.atouife@princeton.edu).

# Data and Code Availability

The code for the optimization model is available at: https://figshare.com/s/a3926a6d8de953d92b55.

# Acknowledgments

This work was funded by Princeton University's Low-Carbon Technology Consortium, which is supported by unrestricted gifts from Google, GE, and ClearPath.

## Author contributions

Conceptualization, M.A. and J.D.J.; methodology, M.A., Y.H and J.D.J; Investigation, M.A.; writing – original draft, M.A. ; writing – review & editing, J.D.J.; funding acquisition, J.D.J.; supervision, J.D.J.

## Declaration of interests

Jesse D. Jenkins is chief scientist of Resilient Transition, which provides modeling, analytics and decision support for utilities, investors, asset owners, and technology companies. He serves on the advisory boards of Eavor Technologies Inc., a closed-loop geothermal technology company, Rondo Energy, a provider of high-temperature thermal energy storage and industrial decarbonization solutions, Dig Energy, a developer of low-cost drilling solutions for geothermal heating and cooling, and Karman Industries, a developer of high-efficiency industrial heat pumps, and he has an equity interest in each company. He also serves as a technical advisor to MUUS Climate Partners and Energy Impact Partners, both investors in early-stage climate technology companies.

## Declaration of generative AI and AI-assisted technologies in the writing process

During the preparation of this work, the authors used ChatGPT and Claude in order to improve readability, language and figures. After using these tools/services, the author reviewed and edited the content as needed and take full responsibility for the content of the publication.

# Figures

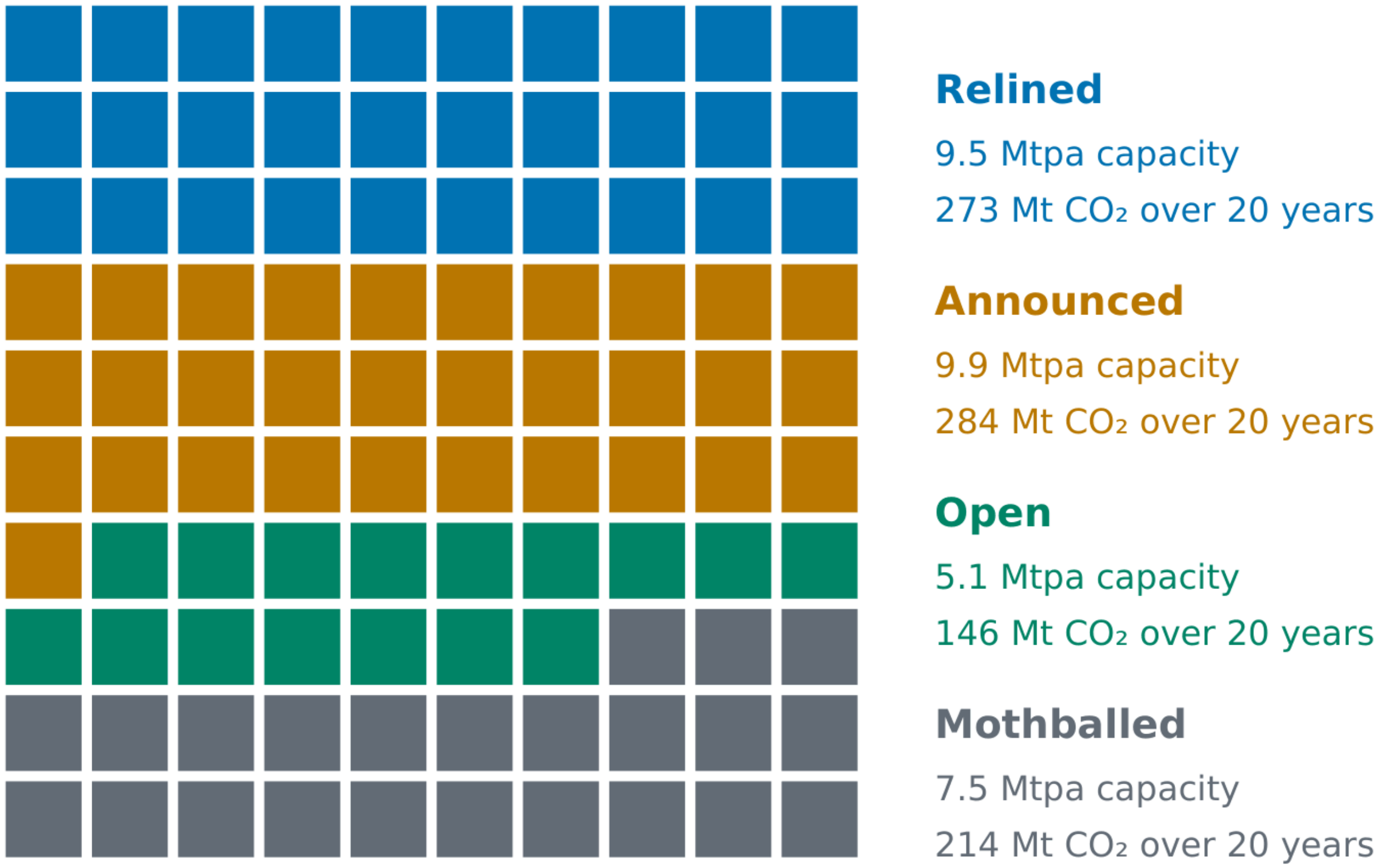

Figure 1: U.S. blast-furnace fleet reinvestment status and 20-year direct $CO_2$ emissions potential.

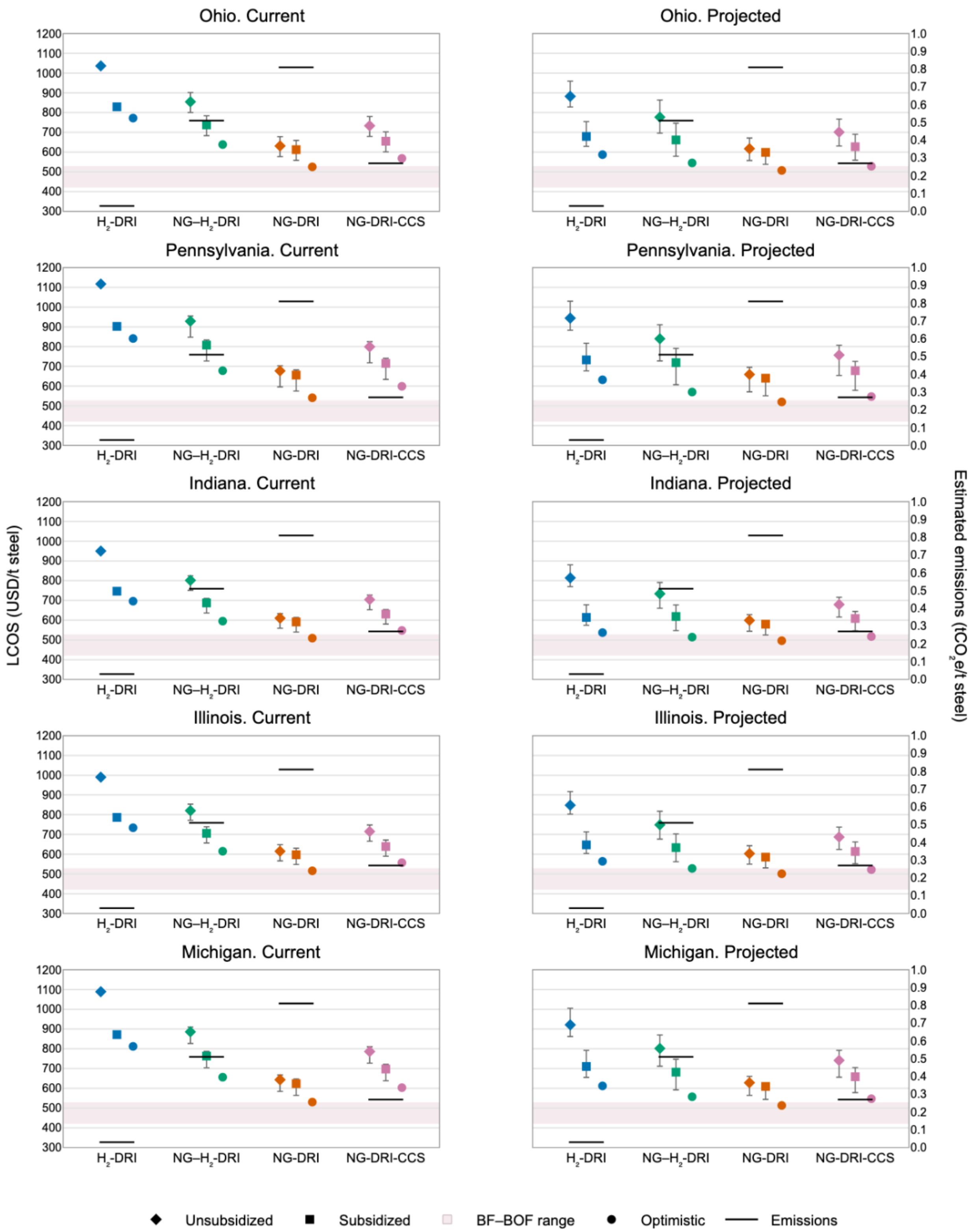

*Figure 2: Current (left) and projected (right) levelized costs of steel production in $ per ton of steel for several low-carbon-emission DRI-EAF processes in states with existing BF-BOF facilities. The shaded area indicates a range of BF-BOF steel production costs.*

*The error bars in the left charts indicate sensitivity to state-specific natural gas prices, whereas the error bars in the right chart indicate sensitivity to projected technology costs (electrolysis, renewables, and storage) and natural gas prices when relevant. The optimistic case is characterized by IRA incentives, a low cost of capital (100 basis points below the base rates), a 30% industrial investment tax credit, and reduced technology costs and $3 per MMBtu natural gas prices when applicable. Projected costs are for the year 2030. The right-side y-axis indicates direct emissions from DRI-EAF processes. Natural gas based processed include contributions of upstream methane leakage at a 2.3% rate (U.S national average).*

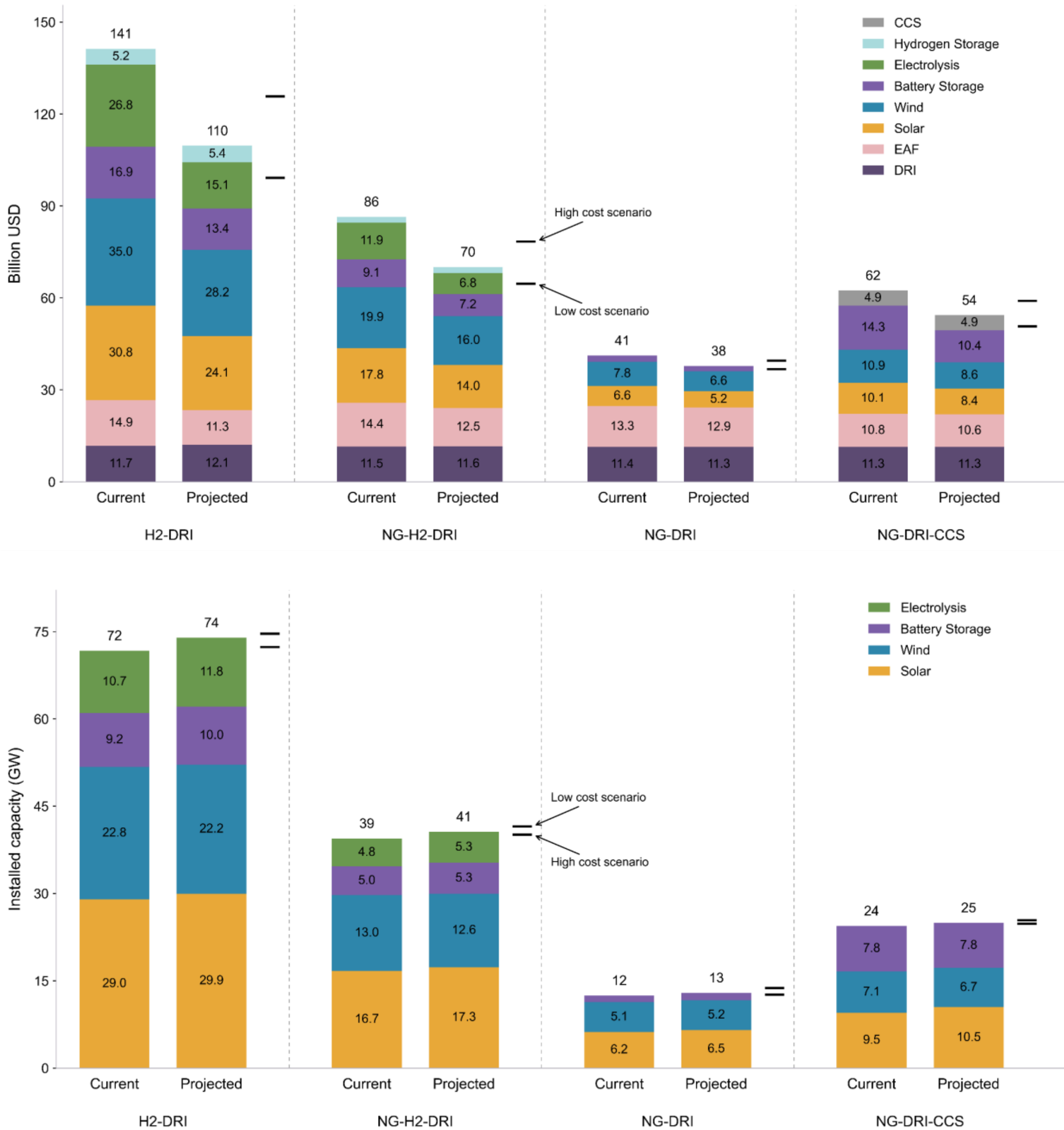

*Figure 3: Current (top) and projected (bottom) total investment and capacity required to transition existing BF-BOF capacity to direct reduction. The error bar indicates uncertainty over future technology costs. "Projected" values are reflected values for projected technology costs in 2030. Both "Current" and "Projected" are based on cases with production and investment tax credits from the Inflation Reduction Act.*

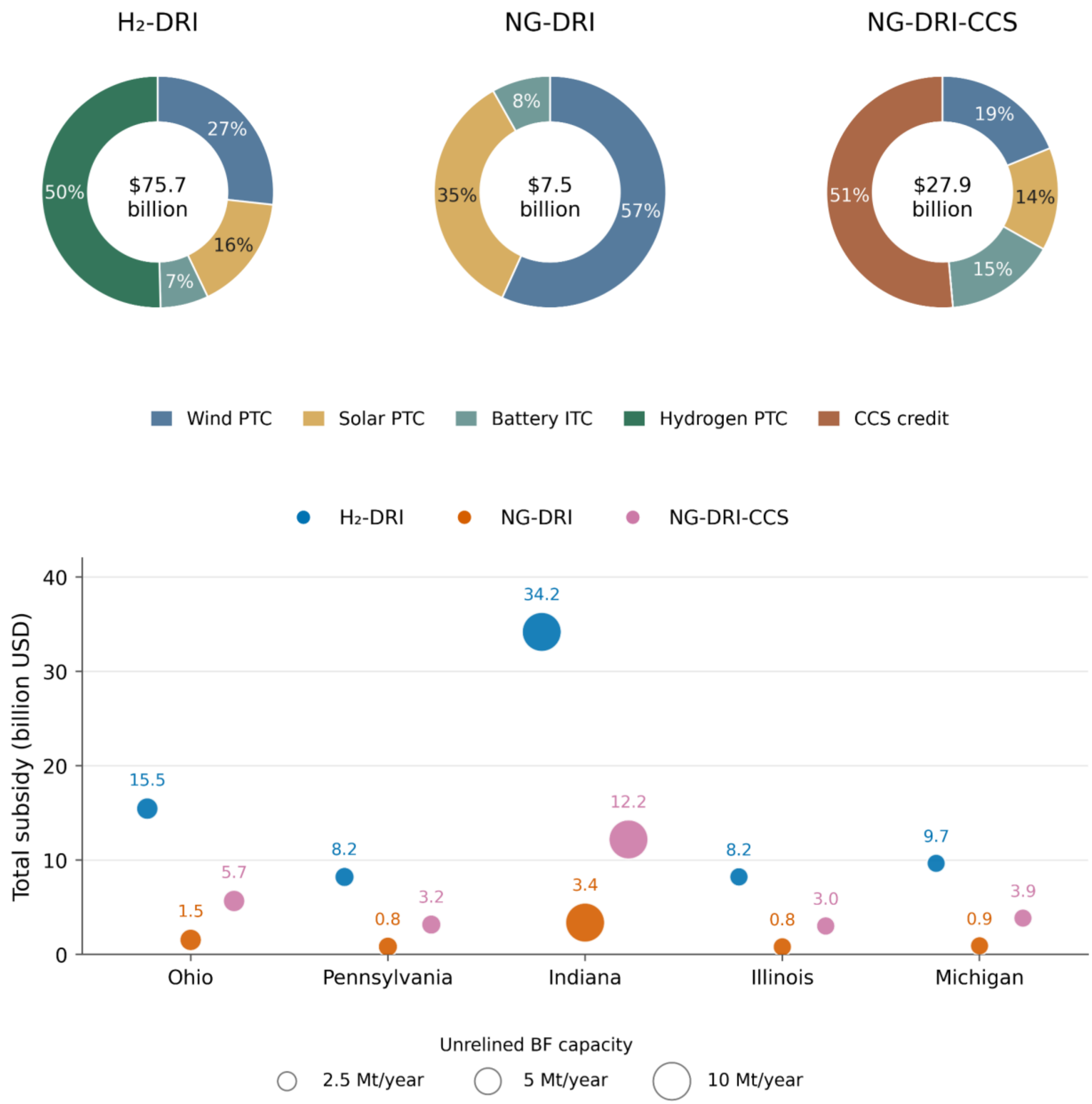

*Figure 4: Inflation Reduction Act subsidies by decarbonization pathway and state. Top: Total subsidies aggregated across all five states by pathway, stacked by credit type. Bottom: Total subsidies by state and pathway; dot size reflects steel-making capacity (mmt/yr).*

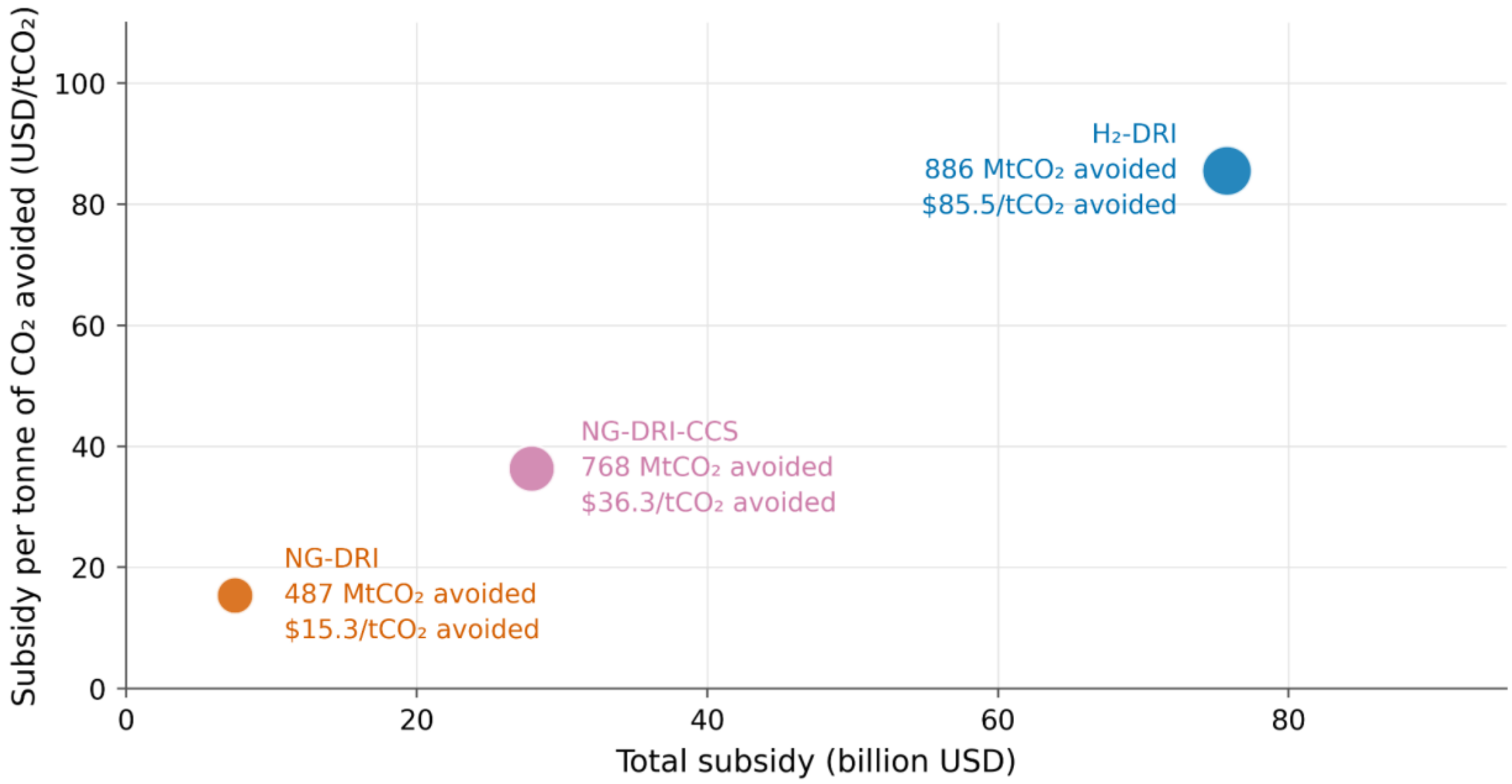

*Figure 5: Total subsidy, total avoided emission, and subsidy per ton of $CO_2$ avoided. Assume all unrelined BF-BOF capacity is converted to specific technology. Total subsidy value assumes Inflation Reduction Act incentives.*